\documentclass[aps,prl,twocolumn,groupedaddress,showpacs]{revtex4-1}

\usepackage{graphicx}

\begin{document}

\title{Absence of supersolidity in solid helium in porous Vycor glass}

\author{Duk Y. Kim}
\author{Moses H. W. Chan}
\email[]{chan@phys.psu.edu}

\affiliation{Department of Physics, The Pennsylvania State University, University Park, PA 16802, USA}

\date{\today}

\begin{abstract}
In 2004, Kim and Chan (KC) carried out torsional oscillator (TO) measurements of solid helium confined in porous Vycor glass and found an abrupt drop in the resonant period below 200 mK. The period drop was interpreted as probable experimental evidence of nonclassical rotational inertia (NCRI). This experiment sparked considerable activities in the studies of superfluidity in solid helium. More recent ultrasound and TO studies, however, found evidence that shear modulus stiffening is responsible for at least a fraction of the period drop found in bulk solid helium samples. The experimental configuration of KC makes it unavoidable to have a small amount of bulk solid inside the torsion cell containing the Vycor disc. We report here the results of a new helium in Vycor experiment with a design that is completely free from any bulk solid shear modulus stiffening effect. We found no measureable period drop that can be attributed to NCRI.
\end{abstract}


\maketitle

In 2004, Kim and Chan (KC) reported torsional oscillator (TO) measurements of solid helium confined in porous Vycor glass \cite{NatureKC}. When the torsion cell was cooled below 200 mK, an abrupt drop in the resonant period of the TO was found. The magnitude of the period drop decreases with the speed of the oscillation of the torsion bob and there is no detectable period drop when the Vycor disc is infused with a solid $^3$He sample. In solid $^4$He samples diluted with $^3$He impurities, the onset temperature of the period drop was found to increase with $^3$He concentration. These observations were interpreted as probable evidence of nonclassical rotational inertia (NCRI) in solid $^4$He. A similar period drop was found in bulk solid $^4$He \cite{ScienceKC}. The NCRI signatures in bulk solid samples show similar dependences on oscillation speed and $^3$He impurities. These results were replicated in over 25 other TO experiments \cite{Shirahama,Reppy07,Kojima07,Kubota,Hunt,Choi,NMRTO,Manchester,Fefferman}.  However, the magnitude of NCRI varies from experiment to experiment, including those carried out in the same laboratory. 

Interestingly, the shear modulus of bulk solid helium showed an increase at the same onset temperature of NCRI with identical temperature and $^3$He concentration dependences \cite{DayBeamish}. For polycrystalline samples a 10 to 20\% increase was found and for single crystal samples, the increase can be as large as 80\% \cite{Rojas10}. Since solid helium is a constituent of the TO, an increase in its shear modulus will stiffen the TO and cause the resonant period to drop, thus mimicking NCRI. This shear modulus effect on the resonant period can be calculated analytically for TOs with simple geometry and also numerically by finite element method (FEM). For an ``infinitely'' rigid TO containing a cylindrical isotropic solid helium sample of 1 cm in diameter and in height and oscillating at 1 kHz, a 20\% increase of the shear modulus of helium results in a drop in the resonant period of approximately $1\times10^{-7}$, or 0.1 ns \cite{Maris11,ReppySM}. For a TO cell that is not rigid, the shear modulus effect can be greatly amplified. In some TOs, solid helium contributes significantly to the mechanical coupling of the different parts of the torsion cell. The coupling is strengthened when the solid helium sample is stiffened at low temperature resulting in a lower resonant period \cite{Reppy07,LongPath}. The torsion rod is usually attached at the center of one end of the (cylindrical) torsion cell. Although the shear modulus of the metal is three orders of magnitude larger ($5\times10^{10}$ versus $2\times10^{7}$ Pa) than solid helium, if the plate to which the torsion rod is attached is sufficiently thin, then solid helium will play a non-negligible role in transmitting the torque from the torsion rod to the rest of the torsion cell. In this case, the resonant period may show a measurable abrupt drop when solid helium is stiffened at low temperature \cite{Maris12}. Lastly, in most TOs torsion rods are hollow and used as fill lines. Solid helium inside the torsion rod contributes to the spring constant of the torsion rod. If the ratio of the outer and inner diameters of the torsion rod is not sufficiently large, then the shear modulus increase at low temperature of the solid helium in the torsion rod can again induce a measurable period drop. It is shown recently that this torsion rod effect may account for the entirety of the apparent NCRI reported in a number of TO experiments \cite{BeamishRod}.

While there is considerable uncertainty on the claim of NCRI in bulk solid helium, at first glance this should not be the case for the 2004 KC Vycor glass experiment. The microscopic mechanism responsible for the shear modulus increase at low temperature in solid helium is the pinning of the dislocation lines to $^3$He impurities at low temperatures. This is not applicable in Vycor glass since the typical distance between neighboring nodes in a dislocation line network is a few microns \cite{Wanner}, orders of magnitude longer than 7 nm, the diameter of the pores in Vycor glass. However, all porous media experiments to date, including KC, were carried out by placing the porous samples inside sealed metal torsion cells \cite{NatureKC,PG,PG2010,Aerogel,Mi}. Since helium must be allowed to infuse into the porous samples, it is necessary to maintain an open empty space, however small, in the torsion cells. When the porous sample is pressurized with solid helium, there will be bulk solid helium in this open space. This raise the question of whether this small quantity of bulk solid in the TO can be responsible for the observed period drop that has been interpreted as the NCRI signature. To clarify this issue, we built a TO out of a Vycor glass disc without a metal container and with a configuration that is completely free from shear modulus stiffening effect of bulk solid helium. Drawings of the current and the KC torsional oscillators are shown in Fig. 1.

\begin{figure}[tb]
   \centerline{\includegraphics[width=1\columnwidth]{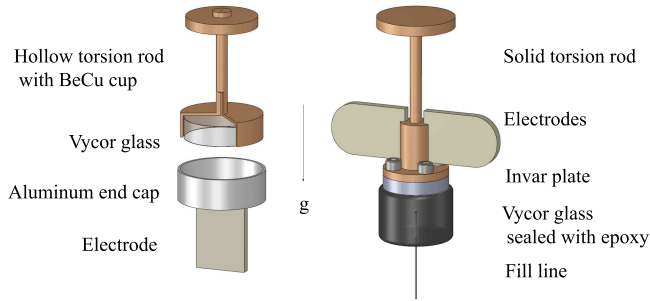}}
   \caption{(Color online) Torsional oscillators with porous Vycor glass used in 2004 (left) and the current experiment (right)}  
   \label{Fig1}
\end{figure}

The Vycor glass disc (diameter 14 mm and height 10 mm) was sealed by a thin layer of epoxy (Stycast2850) painted on the exterior of the disc. A small hole was drilled into the center of the Vycor disc to accommodate a filling capillary (OD 0.3 mm and ID 0.1 mm). The capillary-Vycor joint is also sealed with epoxy. The other end of the capillary is secured in position 6 cm below the Vycor disc. The volume of the empty space inside the Vycor disc surrounding and inside the capillary is estimated to be less than $1\times10^{-4}$ cc, or 0.02\% of the total volume of solid helium inside the porous structure. The expected period changes due to the stiffening of the bulk solid helium inside and outside of the capillary are at most $3\times10^{-5}$ and $5\times10^{-3}$ ns respectively. These values are orders of magnitude smaller than the resolution of the experiment. We know there is no other bulk space in the cell because, if there is any gap between the Vycor disc and the thin epoxy layer, a leak will develop when the sample is pressurized. The Vycor glass disc is glued onto an Invar plate with thermal expansion that matches Vycor. The Vycor-Invar cell was then secured to a solid BeCu torsion rod by screws. The mechanical $Q$ of the empty TO above 0.1 K is $8\times10^{5}$ and as it is commonly found in many other TOs, the $Q$ shoots up below 0.1 K. The resonant period of the TO is 1.15 ms and the mass loading due to a solid helium sample is approximately 5000 ns. Measurements were made on three different liquid helium films adsorbed on the walls of the Vycor pores and superfluid transitions with $T_c$'s of 70, 450 and 630 mK were found. These measurements showed the expected Kosterlitz-Thouless-like behavior confirming that the Vycor disc is free from any crack and capable of supporting superflow. A careful comparison of the temperature dependence of the superfluid density of the lowest coverage film to that of an earlier experiment showed that our temperature scale is reliable to within 1 mK down to 30 mK \cite{Crooker}.

\begin{figure}[tb]
    \centerline{\includegraphics[width=1\columnwidth]{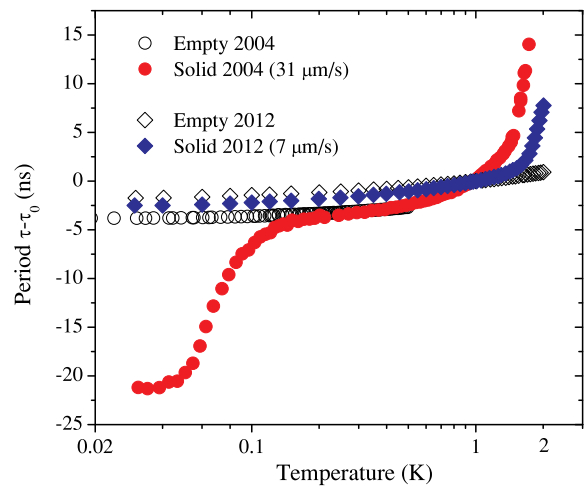}}
    \caption{(Color online) Resonant period vs. temperature of the 2004 (ref. 1) and the 2012 Vycor torsional oscillators. The values of $\tau_0$ of the different data sets are adjusted so that ($\tau-\tau _0$) coincide at 1 K.}
    \label{Fig2}
\end{figure}

The principal results of this experiment are shown in Fig. 2. In contrast to KC, the ``naked'' Vycor TO shows no period drop within experimental resolution (0.1 ns) near and below 200 mK. Since the mass loading is 5000 ns, if there is NCRI, it is less than $2\times10^{-5}$. Both TOs show period drop with decreasing temperature above 1 K. This drop is the signature of solidification of liquid helium in the Vycor pores. In a TO housing a bulk liquid-solid coexistence sample, the resonant period increases when liquid solidifies and adheres to the walls of the torsion cell. However, during the solidification process in Vycor, liquid helium, being entrained in small pores with solid helium ``plugs'', is always coupled to the oscillation. Therefore the resonant period does not increase with solidification but rather decreases because solid helium stiffens the silica structure of the Vycor glass and hence the TO. This is shown more clearly in Fig. 3. The cell was pressurized with 65 bars liquid helium at 2.7 K and cooled down below 1 K rapidly. Then the sample was melted and refrozen a few times with the capillary fill line blocked with solid helium. Figure 3 shows data taken during the fourth warming and cooling cycle and also the fifth warming scan while Fig. 2 shows the data of the same sample during the fourth cooling scan. The thermal cycles show a melting and freezing hysteresis similar to that found in other studies \cite{Molz}. Melting was found to commence near 1.9 K and completed at 2.5 K while freezing is found to commence near 2.0 K and completed near 1.6 K. The period readings show a more rapid change with temperature in the liquid-solid coexistence region than that when the Vycor is filled with either all liquid or all solid. The data taken during the fifth warming scan (from 30 mK) overlap perfectly with the fourth cooling scan up to 1.5 K. When the scan was extended to 2.6 K and cooled down, as in the fourth cycle, the period below 1.5 K were found to be reduced by 1 ns. We think this is a consequence of supercooling of helium in the Vycor pores. As the solid in the Vycor pores is melted when the cell is warmed above 2 K, the liquid immediately refreezes into the open space in and outside the capillary inserted inside the Vycor disc. Thus each warming-cooling cycle up to 2.6 K results in the loss of a small fraction (1 out of 5000) of the solid inside the Vycor pores. The helium from the Vycor pores that accumulated near the capillary at the center of the TO does not contribute any measureable moment of inertia. Other than the 1 ns drop, the period results of the cooling and warming scans between 30 mK and 1.5 K are perfectly reproducible and there is no sign of any abrupt drop within the resolution of the period measurement. The mechanical $Q$ of the TO loaded with a solid sample is 20\% lower than that of the empty cell and there is no sign of any dissipation ``peak'' below 200 mK.

\begin{figure}[tb]
    \centerline{\includegraphics[width=1\columnwidth]{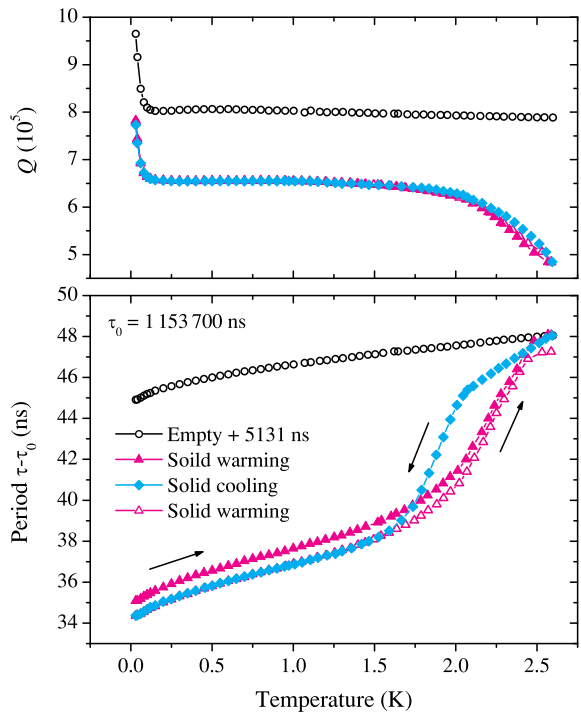}}
    \caption{(Color online) The resonant period and $Q$ of warming and cooling scans of the same solid helium sample shown in Fig. 2. Oscillation speeds are between 7 and 11 $\mu$m/s.}  
    \label{Fig3}
\end{figure}

Figure 4 shows low temperature period and $Q$ readings of two other solid samples, a liquid-solid coexistence sample, an adsorbed (inert layer) film sample at a surface coverage below superfluid onset and solid samples diluted with 30 and 300 ppm of $^3$He. Since the freezing of $^4$He in Vycor takes place over a range of pressure and we have no means of recording the pressure, we labeled each sample in Fig. 4 by the temperature at which freezing is completed. One of the samples we studied shows conversion of liquid to solid down to 1.2 K, where the coexistence boundary is essentially flat. This sample is likely to have a small fraction of liquid down to the lowest temperature. The period versus temperature plots for this liquid-solid coexistence sample and all solid samples including those diluted with $^3$He impurities below 900 mK are reproducible in cooling and warming scans and are perfectly parallel with each other. The temperature dependence of the period is completely insensitive to the oscillation speed (from 1.8 to 78 $\mu$m/s) of the TO. The total change in period between 30 and 900 mK of these plots is 2.5 ns whereas it is 1.6 ns for the empty TO and the inert layer film sample. These observations are consistent with the interpretation mentioned above that the solidification of helium in the porous structure stiffens the Vycor disc and that the stiffness of the Vycor-solid helium composite continue to increase gradually with decreasing temperature down to 30 mK.

\begin{figure}[tb]
    \centerline{\includegraphics[width=1\columnwidth]{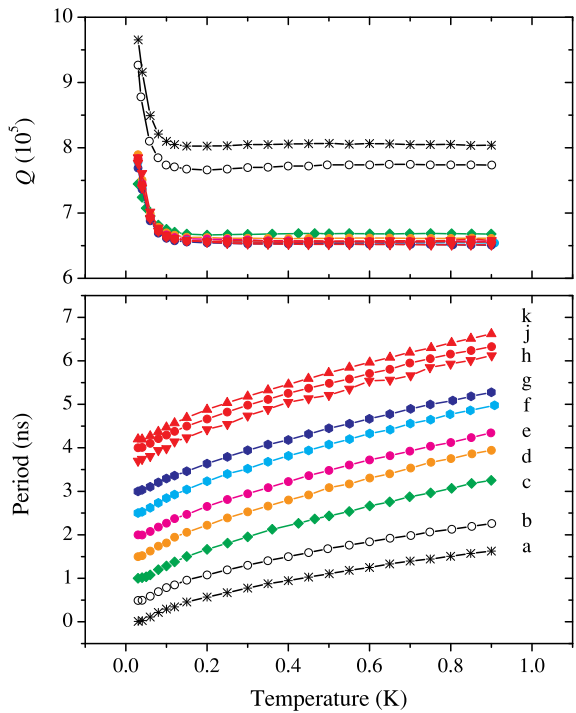}}
    \caption{(Color online) Resonant period and $Q$ of different samples. (a) empty cell, (b) inert layer film (21.5 $\mu$mol/m$^2$), (c) liquid-solid coexistence sample, (d) high purity solid $^4$He (0.3 ppm $^3$He), freezing completed at 1.4 K, (e) high purity solid, freezing completed at 1.9 K, (f) solid $^4$He with 30 ppm $^3$He, freezing completed at 1.5 K, (g) solid $^4$He with 300 ppm $^3$He, freezing completed at 1.6 K, (h, j, k) high purity solid, freezing completed at 1.6 K with oscillation speed of 1.8, 7.3 and 73 $\mu$m/s respectively. Scan j is also shown in Fig. 3. The speeds for all other data sets are between 7 and 11 $\mu$m/s. Period readings are shifted for clarity.}
    \label{Fig4}
\end{figure}

How can we understand the differences in the results of the current and the 2004 KC experiments? We think the period drop found below 200 mK in KC is the result of bulk solid in the metallic torsion cell. There is solid helium in the hollow torsion rod in the TO of KC. The contribution of solid helium towards the spring constant of the torsion rod can be calculated from the shear moduli of BeCu and solid helium and also the OD (2.2 mm) and ID (0.4 mm) of the torsion rod. Such a calculation found that a 20\% increase in the shear modulus of solid helium below 200 mK will result in a period drop of $\sim$0.1 ns. Therefore this is not the primary source of 17 ns drop. In KC, the Vycor disc was first glued with epoxy inside the upper half of the TO, namely an open cylindrical BeCu cup with its top plate connected to the hollow torsion rod. Then an aluminum end cap was slipped over the container and glued with epoxy (Fig. 1). In gluing the Vycor disc, care must be  taken to keep a thin open gap between the top of the disc and the BeCu plate to allow the infusion of helium into the Vycor disc from the torsion rod. This bulk solid helium layer in the gap has a non-negligible contribution to the rigidity of the TO because the thickness of the BeCu plate is only 0.5 mm. FEM simulations show that the effect on the resonant period due to the shear modulus increase is inversely proportional to the thickness of the helium layer. This is sensible since the solid helium layer can be regarded as a glue between the plate of the container and the rest of the TO and it is more ``effective'' if the layer is thin. For a 0.1 mm thick bulk solid helium layer, a 9 ns period change is found for a 20\% increase of shear modulus. For a 50 $\mu$m layer, a period drop of 16 ns, in good agreement with the experimental value of 17 ns, is found. It is also possible that a small gap may open up between the Vycor and the epoxy or between the epoxy and the cylindrical wall of the torsion cell upon cooling and/or pressurization of the TO cell with the helium sample. Solid helium in such a gap will add to the shear modulus stiffening effect. If a bulk solid $^4$He layer is in fact responsible for the apparent NCRI, then it is easy to understand why KC found similar $^3$He impurity and oscillation speed dependences as bulk solid samples. The reason why no period drop was found when the cell was filled with a solid $^3$He sample is that the bulk $^3$He layer in the KC cell has bcc crystal structure which does not show any shear modulus stiffening at low temperature \cite{West}.

Mi and Reppy carried out a recent study of solid $^4$He in Vycor with a TO with two resonant frequencies \cite{Mi}. Similar to KC, the Vycor disc was glued inside an aluminum torsion cell. They found period drops below 200 mK of 1.4 ns in low frequency mode and 0.9 ns in high frequency mode. These period drops are very likely also consequence of the bulk $^4$He solid in the torsion cell and in the hollow torsion rod. The solid helium in the torsion rod can contribute period drops of 0.3 and 0.1 ns in the low and high frequency modes respectively. The effect due to the bulk solid layer in Mi and Reppy’s cell is much smaller than that in KC because the thickness of the plate holding the torsion rod is 10 times thicker. FEM calculations find a solid layer of 0.1 mm will contribute period drops of 0.6 and 0.4 ns respectively in the low and high frequency modes for a 20\% increase in the shear modulus. Since we do not know the actual thicknesses of the bulk helium layers in the torsion cells, the results of the FEM calculations cannot be quantitatively compared with the experimental results. However, the calculations do find period drops in reasonable agreement to both the KC and Mi and Reppy experiments. In the case of the Mi and Reppy experiment, the FEM results also reflect properly the frequency dependence.

In conclusion, we have built a Vycor TO that is free from shear modulus stiffening effect of bulk solid $^4$He and found for solid $^4$He confined in Vycor there is no evidence of any abrupt drop in the resonant period that can be interpreted as NCRI.

\begin{acknowledgments}
We acknowledge useful discussions with John Beamish, Eunseong Kim and John Reppy and we thank Xiao Mi and John Reppy for providing us the dimensions of their TO. Support for this experiment was provided by NSF Grants No. DMR1103159.
\end{acknowledgments}

\bibliography{Vycor2012}

\end{document}